\documentclass[12pt]{article}
\begin{document}
\input epsf
\baselineskip=15pt

\newcommand{\be}{\begin{equation}}
\newcommand{\ee}{\end{equation}}
\newcommand{\bq}{\begin{eqnarray}}

\newcommand{\eq}{\end{eqnarray}}
\newcommand{\x}{{\bf x}}
\newcommand{\y}{{\bf y}}

\newcommand{\p}{\varphi}
\newcommand{\del}{\nabla}
\begin{titlepage}
\vskip1in
\begin{center}
{\Large \bf Order ${\bf 1/N^3}$ corrections to the conformal
anomaly of the (2,0) theory in six dimensions}
\end{center}
\vskip1in
\begin{center}
{\large Paul Mansfield$^a$, David Nolland$^b$ and Tatsuya Ueno$^b$

\vskip20pt

$^a$Department of Mathematical Sciences

University of Durham

South Road

Durham, DH1 3LE, England

{\it P.R.W.Mansfield@durham.ac.uk}

\vskip20pt

$^b$Department of Mathematical Sciences

University of Liverpool

Liverpool, L69 3BX, England

{\it nolland@liv.ac.uk}

{\it ueno@liv.ac.uk} }

\end{center}
\vskip1in
\begin{abstract}
\noindent Using Supergravity on $AdS_7\times S^4$ we calculate the
bulk one-loop contribution to the conformal anomaly of the (2,0)
theory describing $N$ coincident M5 branes. When this is added to
the tree-level result, and an additional subleading order
contribution calculated by Tseytlin, it gives an expression for
the anomaly that interpolates correctly between the large $N$
theory and the free (2,0) tensor theory corresponding to $N=1$.
Thus we can argue that we have identified the exact $N$-dependence
of the anomaly, which may have a simple protected form valid away
from the large $N$ limit.
\end{abstract}

\end{titlepage}


\section{Introduction}

The low energy (2,0) theory describing $N$ coincident M5 branes is
not yet well understood. However some information about this
theory can be obtained via the AdS/CFT correspondence. For
example, the elegant calculation of Henningson and Skenderis
\cite{hs} makes a prediction of the leading order $N$ dependence
of the conformal anomaly of this theory.

One might hope that, as in the case of ${\cal N}=4$ SYM, the
anomaly has a protected form with simple dependence on $N$. Then
provided one could calculate the appropriate sub-leading order
corrections, one would have an exact result that is valid beyond
the large-$N$ regime.

This is indeed what happens for the R-symmetry anomaly of the
(2,0) theory \be J_8=NJ_8^{free}+(N^3-N)p_2,\ee where $J_8^{free}$
is the anomaly of the free ($N=1$) theory and $p_2$ is the normal
bundle of the brane world-volume. As a function of $N$, this
interpolates between the $N=1$ theory and the interacting
large-$N$ one that can be described by 11d Supergravity on
$AdS_7\times S^4$.

The situation is however more complicated than the analogous one
for ${\cal N}=4$ SYM, where a one-loop calculation of the
conformal and R-symmetry anomalies gives $N^2-1$ copies of the
free $N=1$ theory anomalies, and this is protected by
supersymmetry from higher loop and stringy corrections. Take, for
example, the conformal anomaly. It is given in general by a sum of
type-A and type-B anomalies proportional to Euler and Weyl
invariants respectively, and in the ${\cal N}=4$ SYM theory these
are related respectively to three and two point correlators of the
stress tensor. Thus known renormalisation theorems for these
correlators apply.

In general the conformal anomaly of an even-dimensional theory (up
to total derivative terms that are renormalisation scheme
dependent) is a sum of type-A and type-B anomalies, where the
former is proportional to the Euler density and the latter is a
weighted sum of Weyl invariants made out of contractions of the
Weyl tensor and its conformally covariant derivatives. In six
dimensions there are three such Weyl invariants.

For the (2,0) theory the leading order coefficient of the Euler
invariant in the type-A anomaly is related to the four-point
correlator of the stress tensor, while the coefficients of Weyl
invariants in the type-B anomaly depend on the two and three point
correlators. It has been shown \cite{tseytlin2} that the leading
order dependence of the two and three point correlators is given
by $4N^3$ times the corresponding correlators of the free tensor
multiplet. But there is no reason to expect the four-point
correlator to exhibit the same ratio. Indeed, if we look at the
leading order result \cite{hs} for the (2,0) conformal anomaly, we
discover that the type-B anomaly is given by $4N^3$ times that for
the free theory, while the type-A anomalies have a different
ratio, $16N^3/7$ \cite{tseytlin}.

In \cite{d2,d3,d4} we checked the sub-leading order correction to
the conformal anomaly of ${\cal N}=4$ SYM by a one-loop
calculation in $AdS_5\times S^5$ supergravity. In this paper we
will perform a similar calculation on $AdS_7\times S^4$ in order
to calculate sub-leading order corrections to the conformal
anomaly of the large-$N$ (2,0) theory. An attempt to calculate
such corrections by considering $R^4$ corrections to the
supergravity action was made in \cite{tseytlin3}, but our result
gives corrections at a different order in $N$ since $R^4$
corrections give anomalies of $O(N)$, while one-loop supergravity
anomalies are $O(1)$. The different order of these results is
explained by the fact that the supergravity loop-counting
parameter is $G_{Newton}\sim 1/N^3$, whereas the string
loop-counting parameter is $g^2_s\sim 1/N^2$.

Summing over contributions from all the Kaluza-Klein towers of
supergravity fields gives a contribution to the anomaly which,
when properly regularised, is equal to twice the contribution from
a free tensor multiplet. Remarkably, the fields that contribute to
the regularised sum exactly match the field content of the tensor
multiplet; this is similar to what we observed in the $d=4$ case,
where the regularised contributions from Kaluza-Klein fields in
supergravity correspond to a sum of contributions that exactly
match the field content of the ${\cal N}=4$ SYM theory \cite{d4}.

In \cite{tseytlin3} an $O(N)$ contribution to the type-B anomaly
was calculated from $R^4$ terms in the string theory effective
action, but this contribution was conjectured to be
incomplete.(Similar calculations of subleading order anomalies
from $R^4$ terms were performed for the ${\cal N}=4$ SYM case in
\cite{no,blau}.) The contribution calculated in \cite{tseytlin3}
can be seen to be related by supersymmetry to an $O(N)$ term in
the chiral anomaly, but we would expect there to be other
subleading order corrections \cite{moore}. However, any additional
corrections due to stringy effects will not contribute at the same
order as the supergravity contribution.

If we add our $O(1)$ contribution to the type-B anomaly, we get a
result that interpolates correctly between the large-$N$ and $N=1$
cases. Thus our result may give the exact $N$ dependence of the
type-B anomaly. The $O(N)$ contribution to the type-A anomaly was
not calculated in \cite{tseytlin3}, but our results lead us to a
new conjecture for the exact form of the type-A anomaly.

\section{Leading order anomaly from AdS/CFT}

The leading order result of \cite{hs} for the conformal anomaly of
the large-$N$ (2,0) theory can be written as

\be {\cal A}=-{4N^3\over(4\pi)^3\cdot
288}\left[E_6+8(12I_1+3I_2-I_3)+O(\del_iJ^i)\right],\ee

where in terms of the 17 invariants

\bq(A_1,A_2,\ldots,A_{17})&=&(\del^4R,(\del_iR)^2,(\del_iR_{jk})^2,
\del_iR_{jk}\del^jR^{ik},
(\del_iR_{jklm})^2,R\del^2R,R_{ij}\del^2R^{ij},\nonumber\\
&&R_{ij}\del_k\del^jR^{ik},R_{ijkl}\del^2R^{ijkl},R^3,RR_{ij}^2,
RR_{ijkl}^2,R_i^{\
j}R_j^{\ k}R_k^{\ i},R_{ij}R_{kl}R^{ikjl},\nonumber\\
&&R_{ij}R^{iklm}R^j_{\ klm},R_{ij}^{\ \ kl}R_{kl}^{\ \
mn}R_{mn}^{\ \ \ ij},R_{jkjl}R^{imjn}R^{k\ l\ }_{\ m\ n}), \eq

we have \be
E_6=-8A_{10}+96A_{11}-24A_{12}-128A_{13}-192A_{14}+192A_{15}-32A_{16}+64A_{17},
\ee and \be
I_1={19\over800}A_{10}-{57\over160}A_{11}+{3\over40}A_{12}+{7\over16}A_{13}
+{9\over8}A_{14}-{3\over4}A_{15}-A_{17}, \ee \be
I_2={9\over800}A_{10}-{27\over40}A_{11}+{3\over10}A_{12}+{5\over4}A_{13}
+{3\over2}A_{14}-3A_{15}+A_{16}, \ee
\be
I_3=-{11\over50}A_{10}+{27\over10}A_{11}-{6\over5}A_{12}-{3}A_{13}-{4}A_{14}
+{4}A_{15}+{1\over10}A_6-A_7+A_9+\del_i{\cal J}^i . \ee

\section{Seeley-De Witt coefficients}

The conformal anomaly contributed by a conformal field in
six-dimensions is proportional to the Seeley-De Witt coefficient
$b_6$ of the associated kinetic operator. The general expression
for $b_6$ for a six-dimensional operator of the form $-\del^2-E$
was given in \cite{gilkey} and can be written in the form

\bq
b_6&=&{1\over(4\pi)^37!}{\rm tr}\biggl[18A_1+17A_2-2A_3-4A_4+9A_5+28A_6
-8A_7+24A_8+12A_9\nonumber\\
&&+{35\over9}A_{10}-{14\over3}A_{11}+{14\over3}A_{12}-{208\over9}A_{13}
+{64\over3}A_{14}-{16\over3}A_{15}
+{44\over9}A_{16}+{80\over9}A_{17}\nonumber\\
&&+14\biggl(8V_1+2V_2+12V_3-12V_4+6V_5-4V_6+5V_7+6V_8+60V_9+30V_{10}
+60V_{11}\nonumber\\
&&+30V_{12}+10V_{13}+4V_{14}+12V_{15}+30V_{16}+12V_{17}+5V_{18}-2V_{19}
+2V_{20}\biggr)\biggr], \label{b6}\eq
where the invariants $V_a$,
depending on the connection curvature $F_{ij}$ and the
endomorphism $E$, are given by

\bq
(V_1,V_2,\ldots,V_{20})&=&(\del_iF_{jk}\del^iF^{jk},\del^iF_{ji}
\del_kF^{jk},F_{ij}\del^2F^{ij},F_{ij}F^{jk}F_k^{\
i},R_{ijkl}F^{ij}F^{kl},\nonumber\\
&&R_{ij}F^{ik}F^j_{\
k},RF_{ij}F^{ij},\del^4E,E\del^2E,\del_kE\del^kE,E^3,EF_{ij}^2,R\del^2E,
\nonumber\\
&&R_{ij}\del^i\del^jE,\del_iR\del^iE,EER,E\del^2R,ER^2,ER_{ij}^2,ER_{ijkl}^2).
\eq

For a conformally invariant operator, $b_6$ has the general form

\be b_6=aE_6+b_1I_1+b_2I_2+b_3I_3+\del_iJ^i.\ee

The $b_6$ coefficients for the fields appearing in the free (2,0)
tensor multiplet were calculated in \cite{tseytlin}. If we ignore
the total derivative terms, and denote the $b_6$ coefficients for
a conformal scalar, Dirac fermion, and gauge 2-form as $s$, $f$
and $g_{a_2}$ respectively, then we have

\be
s={1\over(4\pi)^37!}\left(-{5\over72}E_6-{28\over3}I_1+{5\over3}I_2+2I_3\right),
\ee

\be
f={1\over(4\pi)^37!}\left(-{191\over72}E_6-{896\over3}I_1-32I_2+40I_3\right),
\ee

\be
g_{a_2}={1\over(4\pi)^37!}\left(-{221\over4}E_6-{8008\over3}I_1
-{2378\over3}I_2+180I_3\right).
\ee

A free (2,0) tensor multiplet consists of 5 scalars, 2 Weyl
fermions and a chiral 2-form gauge field, and its total conformal
anomaly is thus given by

\be 5s+f+g_{a_2}/2=-{1\over(4\pi)^3\cdot288}\left({7\over4}E_6
+8(12I_1-4I_2+I_3)\right). \label{freeanom}
\ee

\section{One-loop conformal anomalies from AdS/CFT}

The one-loop contribution to the conformal anomaly from bulk
supergravity fields was found in \cite{d1} using Schr\"odinger
functional methods. These are particularly appropriate to the
study of the AdS/CFT correspondence, because being Hamiltonian,
they allow us to study bulk fields via sources that live near the
AdS boundary. The result of \cite{d1} can be expressed (for
six-dimensional boundaries) as
\be \delta{\cal
A}=-\sum{(\Delta-3)\over2}b_6, \ee where the sum is taken over all
the fields in 11d Supergravity compactified on $AdS_7\times S^4$,
and $\Delta$ is the scaling dimension of the corresponding
boundary operator.

To find the coefficient $b_6$ appropriate to each bulk field, it
is necessary to decompose the seven-dimensional bulk fields into
ones appropriate to the six-dimensional boundary. There are some
interesting features of this decomposition.

If the boundary is assumed Ricci-flat, then the bulk AdS metric
(satisfying the Einstein equations with cosmological constant
$-15/l^2$) can be written as

\be ds^2={1\over t^2}\left(l^2dt^2+\sum_{i,j}\hat
g_{ij}dx^idx^j\right),\qquad t>0 \label{met}\ee

where $\hat g_{ij}$ is proportional to the boundary metric. In
this metric, the decomposition into boundary fields exhibits
cancellations that ensure that each massive seven-dimensional bulk
field contributes to the anomaly via the Seeley-DeWitt coefficient
corresponding to an irreducible six-dimensional operator with the
same spin. So, for example, the contribution of the massive
seven-dimensional vector field is proportional to the $b_6$
coefficient for the six-dimensional (gauge-fixed) Maxwell
operator. Where there are gauge invariances, there are additional
contributions associated with Faddeev-Popov ghosts.

If the boundary is not Ricci-flat, the metric that satisfies
Einstein's equations is obtained by multiplying $\hat g_{ij}$ in
(\ref{met}) by the factor $(1-\hat Rt^2l^2/120)^2$ where $\hat R$
is the Ricci scalar constructed from $\hat g_{ij}$. The effect of
this on the decomposition into six-dimensional fields is to
introduce couplings to $\hat R$ that render them conformally
covariant. Thus a seven-dimensional minimally coupled scalar
contributes via the $b_6$ coefficient for a six-dimensional
conformal scalar, and a seven-dimensional gauge field via the
$b_6$ coefficient of a six-dimensional gauge field.

Now this necessarily spoils some of the cancellations that we
observed in the Ricci-flat case. For example, by decomposing a
seven-dimensional massive vector into transverse and longitudinal
parts, we can show that the $b_6$ coefficient for it differs from
that of the gauge field by a conformal scalar contribution. In the
Ricci-flat case this cancelled the contribution from the
Faddeev-Popov ghosts, but since the latter are minimally coupled,
the cancellation is now incomplete. However, this is exactly what
is needed to make the overall sum of $b_6$ coefficients a sum of
$b_6$ coefficients of {\em conformal} operators.

In Table 1 we display the values of $\Delta-3$ for the
Kaluza-Klein spectrum. These are related to the bulk masses, which
were first given in \cite{nieu}. The supermultiplets are labelled
by an integer $p\ge2$ and form representations of $USp(4)$. The
$p_6$ coefficients of the fields can be calculated using the
formula (\ref{b6}), but we will not give them all explicitly,
since the only ones we will need in the final result are the ones
involved in the free (2,0) tensor multiplet.

\begin{table}[b]
\begin{center}
\caption{The $(a,b)$ representation of $USp(4)$ has dimension
   $(a+1)(b+1)(a+b+2)(2b+a+3)/6$.}
\label{spec} \vskip .3cm
  \begin{tabular}{|cccc|}
\hline

     Field  & $SU(4)$ rep$^{\rm n}$ & $USp(4)$ rep$^{\rm n}$ &
$\Delta-3$      \\

  \hline

$\phi^{(1)}$ & $(0,0,0)$ & $(0,p)$ & $2p-3$,\quad $p\ge2$ \\
$\psi^{(1)}$ & $(1,0,0)$ & $(1,p-1)$ & $2p-5/2$,\quad $p\ge2$ \\
$A_{\mu\nu\rho}^{(1)}$ & $(2,0,0)$ & $(0,p-1)$ & $2p-2$,\quad
$p\ge2$ \\
\hline $A_\mu^{(1)}$ & $(0,1,0)$ & $(2,p-2)$ & $2p-2$,\quad
$p\ge2$
\\
$\psi_\mu^{(1)}$ & $(1,1,0)$ & $(1,p-2)$ & $2p-3/2$,\quad $p\ge2$
\\
$h_{\mu\nu}$ & $(0,2,0)$ & $(0,p-2)$ & $2p-1$,\quad $p\ge2$ \\
\hline $\psi^{(2)}$ & $(0,0,1)$ & $(3,p-3)$ & $2p-3/2$,\quad
$p\ge3$
\\
$A_{\mu\nu}$ & $(1,0,1)$ & $(2,p-3)$ & $2p-1$,\quad $p\ge3$
\\
$\psi_\mu^{(2)}$ & $(0,1,1)$ & $(1,p-3)$ & $2p-1/2$,\quad $p\ge3$ \\
$A_{\mu\nu\rho}^{(2)}$ & $(0,0,2)$ & $(0,p-3)$ & $2p$,\quad $p\ge3$ \\
\hline
$\phi^{(2)}$ & $(0,0,0)$ & $(4,p-4)$ & $2p-1$,\quad $p\ge4$ \\

$\psi^{(3)}$ & $(1,0,0)$ & $(3,p-4)$ & $2p-1/2$,\quad $p\ge4$
\\
$A_\mu^{(2)}$ & $(0,1,0)$ & $(2,p-4)$ & $2p$,\quad $p\ge4$ \\
$\psi^{(4)}$ & $(0,0,1)$ & $(1,p-4)$ & $2p+1/2$,\quad $p\ge4$ \\
$\phi^{(3)}$ & $(0,0,0)$ & $(0,p-4)$ & $2p+1$,\quad $p\ge4$ \\
\hline
  \end{tabular}
  \end{center}
\end{table}
If we denote the values of $b_6$ for the fields $\phi$, $\psi$,
$A_\mu$, $A_{\mu\nu}$, $A_{\mu\nu\rho}$, $\psi_\mu$, $h_{\mu\nu}$
by $ s,f,v,a_2,a_3,r, $ and $g$ respectively then the contribution
from a generic ($p\ge 4$ ) multiplet is

\bq \left(\sum (\Delta-3)b_6\right)_{p\ge 4}&=&
(-13s+2v-a_2-4f)+(65s+54f-14a_3+6v+2r-g+21a_2){p\over 6}\nonumber\\
&&+(-37s-6f+22a_3+18v+14r+5g-9a_2){p^2\over6} \nonumber \\
&&+(-28s-48f-8a_3-24v-16r-4g-12a_2){p^3\over3}\nonumber\\
&&+(14s+24f+4a_3+12v+8r+2g+6a_2){p^4\over3} \eq

whilst for the $p=3$ multiplet it is \be \left(\sum
(\Delta-3)b_6\right)_{p=3}= 90s+230f+140v+94r+2g+50a_2+62a_3. \ee
The $p=2$ multiplet contains gauge fields requiring the
introduction of Faddeev-Popov ghosts. These are detailed in Table
2, and the total contribution of the $p=2$ multiplet is \be
\left(\sum (\Delta-3)b_6\right)_{p=2}=-16s+10f+16v+10r+3g+10a_3
\ee

Note that if we substitute the values of the $b_6$ coefficients,
the contributions from each supermultiplet are non-zero, even in
the Ricci-flat case (this is unlike the $d=4$ case). To deal with
the sum over multiplets labelled by $p$, we will use the
regularisation introduced in \cite{d4}. The divergent sum is
evaluated by weighting the contribution of each supermultiplet by
$z^p$. The sum can be performed for $|z|<1$ and we take the result
to be a regularisation of the weighted sum for all values of $z$.
Multiplying this by $1/(z-1)$ and integrating around the pole at
$z=1$ gives a regularisation of the original divergent sum.

This yields \be \sum (\Delta-3)b_6=26s+4f-4v+a_2. \ee As discussed
earlier, the $b_6$ coefficients of massive fields depend on the
decomposition from seven to six dimensions. For the massive
vector, we have \be 2v=2v_0+3s-3s_0, \ee where $v_0,s_0$ are the
coefficients for the gauge-fixed six-dimensional Maxwell operator
and minimally coupled scalar, respectively \cite{d5}. The $b_6$
coefficients for all other massive fields are what we would expect
for the appropriate spin, for example the contribution for the
massive graviton corresponds to the heat-kernel coefficient for
the transverse traceless part of a six dimensional spin 2
operator, and the contribution for a two-index antisymmetric
tensor is the heat-kernel coefficient for an irreducible six
dimensional operator of the same spin.

The final expression for the one-loop contribution to the
conformal anomaly is  \be \delta {\cal A}=-\sum
{(\Delta-3)b_{6}/2}=-2(5s+f+g_{a_2}). \label{afsum} \ee

\section{Discussion}

If we express the result (\ref{afsum}) in terms of the Euler and
Weyl invariants we get

\be \delta {\cal
A}={1\over(4\pi)^3\cdot288}\left({7\over2}E_6+16(12I_1+3I_2-I_3)\right),
\label{sub1}\ee

which is to be added to the leading order result

\be {\cal A}=-{4N^3\over(4\pi)^3\cdot
288}\left[E_6+8(12I_1+3I_2-I_3)+O(\del_iJ^i)\right].\ee

In \cite{tseytlin3} an additional subleading order contribution to
the anomaly was identified. Since the topology of the boundary was
assumed to be trivial, implying the vanishing of the Euler
density, only the contribution to the coefficients of the Weyl
invariants was determined. This is given by

\be \delta {\cal
A}={N\over(4\pi)^3\cdot288}\left(8(12I_1+3I_2-I_3)\right).
\label{sub2}\ee

We can speculate that there is a similar contribution proportional
to the Euler density with an undetermined coefficient $\alpha$:

\be \delta {\cal A}={N\over(4\pi)^3\cdot288}\alpha E_6.
\label{sub3}\ee

Adding all the contributions together gives

\be {\cal A}=-{1\over(4\pi)^3\cdot 288}\left[(4N^3-\alpha
N-{7\over2}) E_6+(4N^3-N-2)\cdot
8(12I_1+3I_2-I_3)+O(\del_iJ^i)\right].\label{total}\ee

Putting $N=1$, we observe that the coefficient of the Weyl
invariants coincides with the result (\ref{freeanom}) for the free
(2,0) tensor multiplet. If $\alpha=-5/4$, the coefficient of the
Euler density would coincide as well. Thus we conjecture that
there is an $O(N)$ contribution to the conformal anomaly
corresponding to (\ref{sub3}) with $\alpha=-5/4$, and that the
exact $N$-dependence of the conformal anomaly is thus

\be {\cal A}=-{1\over(4\pi)^3\cdot
288}\left[(4N^3+{5\over4}N-{7\over2}) E_6+(4N^3-N-2)\cdot
8(12I_1+3I_2-I_3)+O(\del_iJ^i)\right].\ee

\begin{table}[t]
\begin{center}
\caption{Decomposition of gauge fields for the massless
multiplet.} \label{ghosts} \vskip .3cm
\begin{tabular}{|c|cc|}
\hline
Original field & Gauge fixed fields & $\Delta-3$\\
\hline
$A_\mu$ & $A_i$ & 2 \\
       & $A_0$ & 3\\
         & $b_{FP}$, $c_{FP}$ & 3\\
\hline
$\psi_\mu$ & $\psi_i^{\rm irr}$ & 5/2\\
&            $\gamma^i\psi_i$ & 7/2\\
  & $\psi_0$ & 7/2 \\
& $\lambda_{FP}$, $\rho_{FP}$ & 7/2 \\
& $\sigma_{GF}$ & 7/2\\
\hline
$h_{\mu\nu}$ & $h_{ij}^{\rm irr}$ & 3 \\
 & $h_{0i}$ & 4\\
& $h_{00}$, $h_\mu^\mu$ & $\sqrt{18}$\\
& $B^{FP}_0$,$C^{FP}_0$ &$\sqrt{18}$\\
& $B^{FP}_i$,$C^{FP}_i$ & 4\\
\hline
   \end{tabular}
  \end{center}
\end{table}

\end{document}